%% file: belle2.tex
\documentclass[%
 reprint, 
 amsmath,amssymb,
 aps,
    nolongbibliography,
		citeautoscriptwocolumn,
]{revtex4-2}

\usepackage{graphicx} 
\usepackage{dcolumn} 
\usepackage{colordvi}
\usepackage{color}
\usepackage{epstopdf}
\usepackage{amssymb}
\usepackage{url}
\graphicspath{{ps}}
\usepackage{hyperref}
\usepackage{tabularx}
\usepackage{verbatim}
\usepackage{multirow}
\usepackage{siunitx}
\usepackage{orcidlink} 

\input belle2sym.tex

\def\BFCentValue{\ensuremath{{0.939}}\xspace}
\def\BFStatErr{\ensuremath{{0.021}}\xspace}
\def\BFSystErr{\ensuremath{{0.050}}\xspace}

\def\BDrho{\ensuremath{B^- \to D^0 \rho^-}\xspace}
\def\BDpipi{\ensuremath{B^- \to D^0 \pi^-\pi^0}\xspace}
\def\BDpiKpipiz{\ensuremath{B^- \to D^0(\to K^-\pi^+\pi^0) \pi^-}\xspace} 
\def\BDpiKpi{\ensuremath{B^- \to D^0(\to K^-\pi^+) \pi^-}\xspace} 

\begin{document}

\def\belletwo {\it {Belle II}}
\newcommand{\detailtexcount}[1]{%
 \immediate\write18{texcount -merge -sum -q #1.tex output.bbl > #1.wcdetail }%
 \verbatiminput{#1.wcdetail}%
}

\newcommand{\quickwordcount}[1]{%
 \immediate\write18{texcount -1 -sum -merge -q #1.tex output.bbl > #1-words.sum }%
 \input{#1-words.sum} words%
}

\newcommand{\quickcharcount}[1]{%
 \immediate\write18{texcount -1 -sum -merge -char -q #1.tex output.bbl > #1-chars.sum }%
 \input{#1-chars.sum} characters (not including spaces)%
}



\title{Measurement of the branching fraction of the decay \boldmath{$B^- \to D^0 \rho(770)^-$} at Belle II}


\input{authors}

\begin{abstract}
We measure the branching fraction of the decay $B^- \to D^0 \rho(770)^-$ using data collected with the Belle~II detector. The data contain 387 million $\BB$ pairs produced in $e^+e^-$ collisions at  the $\FourS$ resonance.  We reconstruct $8360\pm 180$ decays from an analysis of the distributions of the $B^-$ energy and the $\rho(770)^-$ helicity angle. We determine the branching fraction to be $(\BFCentValue \pm \BFStatErr\stat \pm \BFSystErr\syst)\%$, in agreement with previous results. Our measurement improves the relative precision of the world average by more than a factor of two. 
\end{abstract}

\maketitle

{\renewcommand{\thefootnote}{\fnsymbol{footnote}}}
\setcounter{footnote}{0}

The Belle~II experiment uses \epem collisions at center-of-mass energies of about 10.6\gev to access an abundant sample of \BB pairs produced through $\FourS$ decays and search for nonstandard-model physics in weak decays of $B$ mesons~\cite{Aushev:2010bq,Belle-II:2022cgf}. A large part of this physics program depends on reconstructing the decays of both $B$ mesons: one is the signal, the other (tag decay) is used to infer signal properties or to suppress background from continuum production of light quarks. This analysis technique is known as tagging. Belle~II tagging algorithms use multivariate classifiers trained with simulated events~\cite{Keck:2018lcd}. The algorithms are calibrated on control data to correct for simulation mismodeling, which originates mainly from inaccurate or missing information on branching fractions and decay models of some tag decays~\cite{Belle-II:2020fst}. Improved knowledge of these decays would yield better training, resulting in enhanced and more reliable tagging performance. This will improve Belle~II physics reach. 

Decays into fully hadronic final states offer the best signal-to-background ratio for tagging. The dominant hadronic tag channels are Cabibbo-favored decays into a charm meson and several pions. Among these, \mbox{$B^- \to D^0 \rho(770)^-$} is one of the most effective (charge conjugation is implied throughout the letter unless otherwise stated).  However, the tagging efficiency  differs significantly between data and simulation for this tag decay, strongly suggesting the need to revisit the measurement of its branching fraction. 
The current world average, $(1.34 \pm 0.18)\%$~\cite{PDG_2022}, is dominated by a single measurement from 1994 by the CLEO collaboration, which employed a sample of $e^+e^-$ collisions at the \FourS resonance with an integrated luminosity of $0.9\invfb$~\cite{CLEO:1994mwq}. We note that the world average from Ref.~\cite{PDG_2022} is not updated to the latest branching fractions of $D^0$ and  $\FourS$ decays; scaling them to the current values, the average would be about 10\% smaller. 

Branching-fraction measurements of the decays $B^- \to D^0 \rho^-$, $\Bbar^0 \to D^+ \rho^-$, and $\Bbar^0 \to D^0 \rho^0$,  also  provide tests of calculations of hadronic decay rates based on the heavy-quark limit and factorization models~\cite{Beneke:2000ry,Cheng:1998kd}. Hereafter, we refer to $\rho(770)$ mesons as $\rho$. Using isospin symmetry, their decay amplitudes  can be expressed in terms of amplitudes for  two isospin eigenstates, $A_{1/2}$ and $A_{3/2}$~\cite{Neubert:2001sj,Chua:2001br,Mantry:2003uz,Keum:2003js,BaBar:2011qjw,Rosner:1999zm}. The ratio $R$ and strong-phase difference $\delta$ between these amplitudes are related to the branching fractions $\mathcal{B}(D\rho)$ of the three decays  and the ratio $\tau_+/\tau_0$ of the $B^+$ and $B^0$ lifetimes,  as

\begin{align}
        R &= \left( \frac{3}{2} \frac{\tau_+}{\tau_0} \frac{\mathcal{B}( D^+ \rho^-) + \mathcal{B}(D^0 \rho^0)}{\mathcal{B}( D^0 \rho^-)} - \frac{1}{2} \right)^{\frac{1}{2}}\,, \label{eq:R}\\
        \cos\delta &= \frac{1}{2R} \left( \frac{3}{2} \frac{\tau_+}{\tau_0} \frac{\mathcal{B}( D^+ \rho^-)-2\mathcal{B}(D^0 \rho^0)}{\mathcal{B}( D^0 \rho^-)} + \frac{1}{2} \right) \,. \label{eq:delta}
\end{align}
In the heavy-quark limit, factorization models predict $R = 1 + \mathcal{O}(\Lambda_{\rm QCD}/m_b)$ and $\delta = \mathcal{O}(\Lambda_{\rm QCD}/m_b)$, where $m_b$ is the $b$ quark mass and $\Lambda_{\rm QCD}$ the QCD scale. The LHCb collaboration reported $R = 0.69 \pm 0.15$ and $\cos\delta = 0.984^{+0.113}_{-0.048}$~\cite{LHCb:2015klp}. The accuracy of the test is limited by the 13\% and 16\% fractional uncertainties on the branching fractions of the $B^- \to D^0 \rho^-$ and  $\Bbar^0 \to D^+ \rho^-$ decays, respectively~\cite{PDG_2022}. 

In this letter, we report an improved measurement of the  $B^- \to D^0 \rho^-$ branching fraction using data collected at Belle~II.   We reconstruct the $D^0$ meson in its two-body favored decay $D^0\to K^-\pi^+$ and restrict the $\pi^-\pi^0$ invariant mass to a 300\mevcc range centered at the $\rho^-$ mass pole. In this range, approximately twice the $\rho^-$ natural width, we expect the \BDrho decay to nearly saturate the $D^0 \pi^-\pi^0$ final state, with only a small contribution from the three-body \BDpipi decay~\cite{CLEO:1994mwq}. To determine the \BDrho yield, we analyze the background-subtracted distribution of the helicity angle of the $\rho^- \to \pi^-\pi^0$ decay. The latter is obtained by fitting the $B^-$ energy distribution in nine independent intervals of the helicity angle. 
The \BDrho yield is divided by the reconstruction efficiency and the number of $B^-$ mesons produced to determine the branching fraction. The data sample has an integrated luminosity of 362\invfb, which enables a more precise measurement than the current world average. The main uncertainty is systematic and related to the calculation of the signal efficiency, and to the fit model. To reduce the systematic uncertainty, the selection is chosen to minimize background contamination while retaining a few-percent fractional statistical precision on the branching-fraction measurement. We designed the analysis using simulated and experimental control data before examining the signal region.


The Belle~II experiment~\cite{Abe:2010gxa} is located at SuperKEKB,
which collides electrons and positrons at and near the $\Upsilon(4S)$
resonance~\cite{Akai:2018mbz}. The Belle II
detector~\cite{Abe:2010gxa} has a cylindrical geometry and includes a
two-layer silicon-pixel detector surrounded by a four-layer
double-sided silicon-strip detector 
and a 56-layer central drift chamber. These detectors
reconstruct tracks of charged particles.  Only one sixth of the second
layer of the silicon-pixel detector was installed for the data analyzed here. The
symmetry axis of these detectors, defined as the $z$ axis, is almost
coincident with the direction of the electron beam.  Surrounding the
drift chamber, which also provides \dedx energy-loss measurements, is a
time-of-propagation counter 
in the central region and an aerogel-based ring-imaging Cherenkov
counter in the forward region.  These detectors provide
charged-particle identification.  Surrounding them is an
electromagnetic calorimeter~(ECL) based on CsI(Tl) crystals that
primarily provides energy and timing measurements for photons and
electrons. Outside of the ECL is a superconducting solenoid
magnet. Its flux return is instrumented with resistive-plate chambers
and plastic scintillator modules to detect muons, $K^0_L$ mesons, and
neutrons. The solenoid magnet provides a 1.5~T magnetic field 
parallel to the $z$ axis.

Large samples of simulated data are generated by modeling the physics processes resulting from $\epem$ collisions and propagating the final-state particles through a detailed simulation of the detector.
We use the  \texttt{EVTGEN}~\cite{Lange:2001uf}, \texttt{PYTHIA8}~\cite{Sjostrand:2014zea}, and \texttt{KKMC}~\cite{Jadach:1999vf} software libraries to model particle production and decay, \texttt{PHOTOS}~\cite{Barberio:1993qi} for photon radiation, and \texttt{GEANT4}~\cite{GEANT4:2002zbu} for material interaction and detector response.
The simulation includes beam-induced backgrounds~\cite{Liptak_2022}.
Collision and simulation data are processed using the \texttt{BASF2}~\cite{Kuhr:2018lps, Basf2-zenodo} software. 

The event selection starts online with criteria based on the total energy and charged-particle multiplicity, which are fully efficient for signal and
strongly suppress low-multiplicity events. Offline, we select tracks with loose requirements on their radial ($\delta r<\SI{0.5}{cm}$) and longitudinal ($|\delta z|<\SI{3.0}{cm}$) displacements from the average $e^+e^-$ interaction point. 
We require tracks to be in the polar-angle acceptance of the drift chamber ($\SI{17}{\degree}<\theta<\SI{150}{\degree}$).  

Photons are reconstructed from ECL energy clusters that are not matched to tracks. Photons reconstructed in the central region of the ECL ($\SI{32.2}{\degree}<\theta<\SI{128.7}{\degree}$) are required to have energies greater than 50\,MeV, and those in the forward ($\SI{12.4}{\degree}<\theta<\SI{31.4}{\degree}$) and backward ($\SI{130.7}{\degree}<\theta<\SI{155.1}{\degree}$) end caps greater than 60 and 100\,MeV, respectively. 
To suppress beam background, photon clusters must include more than one ECL crystal and have a signal time within 200\,ns of the collision time. We reconstruct $\piz$ candidates by combining pairs of photons. 
We require the angle between the two photons to be smaller than 1\,rad, the difference in their azimuthal angles to be smaller than 2.2\,rad, and the diphoton mass to be consistent with the known $\pi^0$ mass within approximately two times the resolution. We also exclude extreme values of the cosine of the $\pi^0$ helicity angle. In the $\pi^0$ rest frame, this is the angle between the photon direction and the $\pi^0$ boost direction from the lab frame. We remove candidates that have cosine values larger than 0.98 to suppress combinatorial background from collinear low-momentum photons.
We train a boosted-decision-tree (BDT) classifier with 14 variables associated with cluster shapes to suppress photons misreconstructed from hadronic clusters and hadronic-shower splitting~\cite{Belle-II:2022ihd}. We choose a threshold on the BDT output optimized to select $\pi^0$ candidates from $\rho^-\to \pi^0 \pi^-$ decays.  A requirement that the $\pi^0$ momentum be larger than 240\mevc suppresses low-energy $\pi^0$ candidates from decays of excited charm states. 

Tracks are assumed to be charged pions. We combine a neutral and a charged pion to form a $\rho^-$ candidate. 
We reconstruct $D^0$ candidates by combining oppositely charged kaon and pion candidates with invariant masses between 1.85 and 1.88\gevcc, a range about six units of mass resolution wide. The kaon candidate is a track that satisfies a threshold on the ratio $\mathcal{L}_K/(\mathcal{L}_\pi + \mathcal{L}_K)$, where the likelihood $\mathcal{L}_{\pi,K}$ for a pion or kaon hypothesis combines particle-identification information from all subdetectors except the pixel detector. The requirement retains 95\% of kaons and rejects about 90\% of pions misidentified as kaons. 
We combine $D^0$ and $\rho^-$ candidates through a kinematic fit of the entire decay chain to form $B^- \to D^0 \rho^-$ decays, constraining the $B^-$ to originate from the \epem interaction region and the \piz mass to its known value to improve $B$ energy and momentum resolutions. For each $B$ candidate, we calculate the beam-constrained mass, $M_{\rm bc} = \sqrt{E_{\rm beam}^{* 2}/c^4 - |\Vec{p}_B^{\ *}|^{2} /c^2}$, and the energy difference, $\Delta E = E_B^* - E_{\rm beam}^{*}$, where $E_{\rm beam}^{*}$ is the beam energy, $E_B^*$ is the $B$ energy, and $\Vec{p}_B^{\ *}$ its momentum vector, all computed in the center-of-mass frame. We require $M_{\rm bc}>5.27\gevcc$ and $-0.18 < \Delta E <  0.2$\,GeV. 

We define the $\rho$ helicity angle $\theta_{\rho}$ as the angle between the $\pi^-$ momentum and the direction opposite to the $B^-$ momentum in the $\rho^-$ rest frame. We select the region $\cos\theta_\rho<0.7$, which excludes events with invariant masses $m(D^0\pi^0)$ smaller than  about $2.6$\gevcc. This  removes background from $B$ decays into $D^0 \pi^-\pi^0$ through $D_0^{*0} \to D^0 \pi^0$, $D_2^{*0}\to D^0 \pi^0$, 
which could interfere with the signal. Rates for these processes are not known, but simulation studies show that our selection suppresses to negligible levels any excited $D^*$ contribution regardless of the unknown decay amplitudes. 

To further suppress background, we use all tracks not associated with the signal to reconstruct the decay vertex of the other $B$ meson and to identify its flavor with a dedicated algorithm~\cite{Belle-II:2021zvj}. This information, along with event-shape variables (modified Fox-Wolfram moments~\cite{fw,Belle:ksfw}, energy flows~\cite{CLEO:1995rok}, sphericity-related quantities~\cite{sphericity}, and thrust-related quantities~\cite{thrust_related}), is used as input to a BDT classifier. The BDT is trained with simulation to distinguish signal from continuum background: $e^+e^- \to q\bar{q}$ processes, where $q = u, d, s, c$. The most discriminating inputs are the ratio between the second and zeroth Fox-Wolfram moments, the magnitude of the signal $B$ thrust, and the cosine of the angle between the thrust axis of the signal $B$ and that of all the remaining charged and neutral particles in the event. 
We cut on the BDT output to reduce the continuum background to a negligible level. There are multiple candidates in 4\% of events (2\% in the signal region $|\Delta E|<50$\mev) and are all retained. 

After the selection, simulation shows that $64\%$ of the total background is due to $B$ decays other than signal, the majority being Cabibbo-favored semileptonic decays. 
We  refer to these decays as \BB background. Studies of simulated samples show that backgrounds from suppressed charmless decays, such as $B^- \to K^- \pi^+ \rho^-$ and $B^- \to K^- \pi^+ \pi^- \pi^0$, are negligible. The remaining $36\%$ originates from misreconstructed signal decays (self-cross-feed), where one of the final-state particles is mistakenly taken from the decay of the accompanying $B$ meson. Such misreconstruction only occurs for the $\rho^- \to \pi^-\pi^0$ decay. The neutral pion is misreconstructed using a photon not belonging to the signal in half of the cases of self-cross-feed, and a charged or neutral pion from the other $B$-meson decay is used in the other half.

We determine the sample composition with a maximum likelihood fit to the unbinned distribution of $\Delta E$, including three components: signal, \BB background, and self-cross-feed. At this stage, we treat possible contamination of the three-body decay \BDpipi, where the $\pi^-\pi^0$ system is nonresonant, as part of the signal, as its $\Delta E$ distribution is indistinguishable from that of \BDrho decay. The  $\Delta E$ fit is performed in nine independent intervals of $\cos\theta_{\rho}$ to reconstruct the background-subtracted distribution of this variable and separate \BDrho from \BDpipi decays. As the $\rho^-$ is fully longitudinally polarized in \BDrho decays, 
the helicity angle distribution is proportional to $\cos^2\theta_{\rho}$; by contrast, in nonresonant \BDpipi decays  the $\pi^-\pi^0$ system is in $S$-wave with a uniform $\cos\theta_{\rho}$ distribution. Interference between the amplitudes of the two decays modifies these distributions. 
To maximize the sensitivity at $\cos\theta_{\rho} \simeq 0$, where \BDrho decays are suppressed, we use a variable interval width so that the expected \BDrho distribution is remapped into a uniform distribution. With this  choice, the  \BDpipi  shape peaks in the central region. The interval scheme is chosen by using simulated data and accounts for the efficiency variation as a function of  $\cos\theta_{\rho}$. We require  the same number of \BDrho candidates in each of the nine intervals and we check that the $\cos\theta_\rho$ resolution is negligible compared to the interval width.

\begin{figure}[tb]
    \centering
    \includegraphics[width=0.80\linewidth]{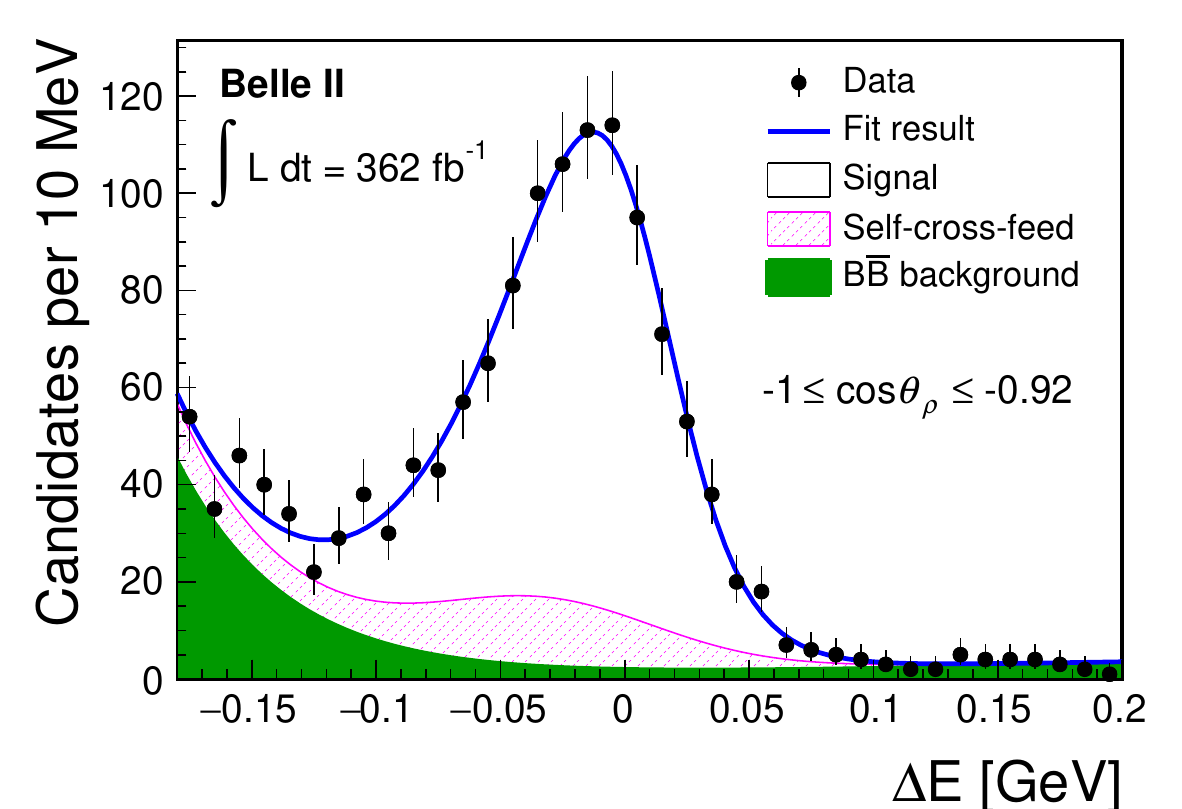}
    \includegraphics[width=0.80\linewidth]{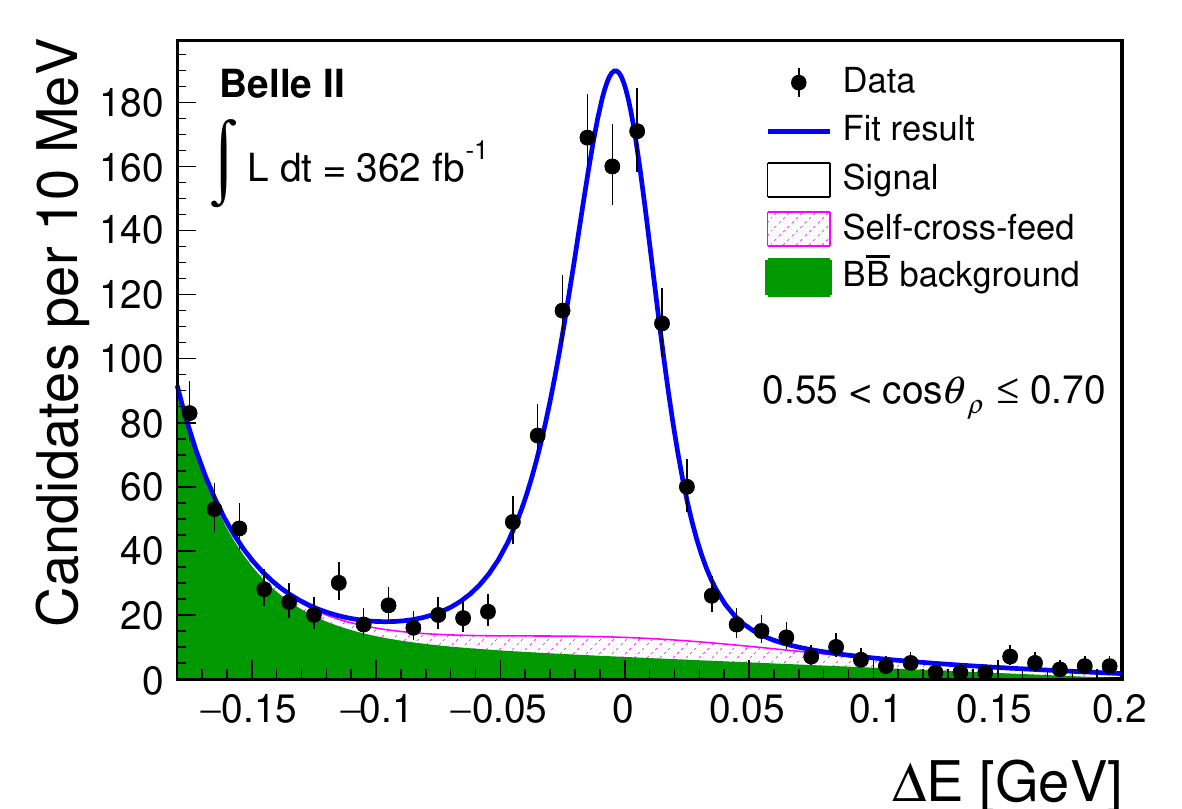}
  \caption{Distribution of $\Delta E$ with fit projection overlaid for (top) the lowest and (bottom) the highest $\cos\theta_{\rho}$ intervals.}
    \label{fig:fit-DeltaE}
\end{figure}

The $\Delta E$ distribution changes significantly across the $\cos\theta_{\rho}$ intervals for each fit component. Examples in the two extremal intervals are shown in Fig.~\ref{fig:fit-DeltaE}. The signal is modeled with a Johnson $S_U$ function~\cite{johnson}, which peaks close to zero in each interval and has a tail at negative $\Delta E$. However, variations of the peak position, width, and skewness are observed  because the energy resolution depends on the momenta of the pions in the $\rho^-$ decay. 
When $\cos\theta_{\rho}$ approaches $-1$, the energy resolution is dominated by the ECL energy resolution on the fast \piz; for $\cos\theta_{\rho}$ values closer to $1$, the largest contribution to the energy resolution comes from the measurement of the momentum of the fast \pim. Self-cross-feed also peaks around zero, but with a broad structure that is modeled by a Gaussian function; in the extremal $\cos\theta_{\rho}$ intervals, an exponential function is also added. The shape of the \BB background is exponential in each $\cos\theta_{\rho}$ interval and peaks at the lower edge of the $\Delta E$ range with different slopes; intervals closer to $\cos\theta_{\rho}=-1$ have a smaller slope.  
All shapes are determined by fitting  simulated data and allow for adjustments across the  $\cos\theta_{\rho}$ intervals. For each interval, the free parameters of the fits to the experimental data are the yields of the signal and \BB background, the exponential slope of the \BB background, and shift- and width-correction parameters common to the signal and self-cross-feed peaks. The ratio of self-cross-feed to signal is also a fit parameter that is Gaussian constrained to the value found in simulation; the uncertainty on this constraint is discussed below.
The $\Delta E$ distribution is modeled well in
each $\cos\theta_{\rho}$ interval, with $p$ values from 0.07 to 0.96. The $p$ value of a fit is calculated as the $\chi^2$ probability from the pull distribution of the fit projection with 26 bins in $-0.18 < \Delta E < 0.08$\,GeV, with degrees of freedom equaling the number of bins minus the free parameters of the fit. 

\begin{figure}[tb]
    \centering
    \includegraphics[width=0.80\linewidth]{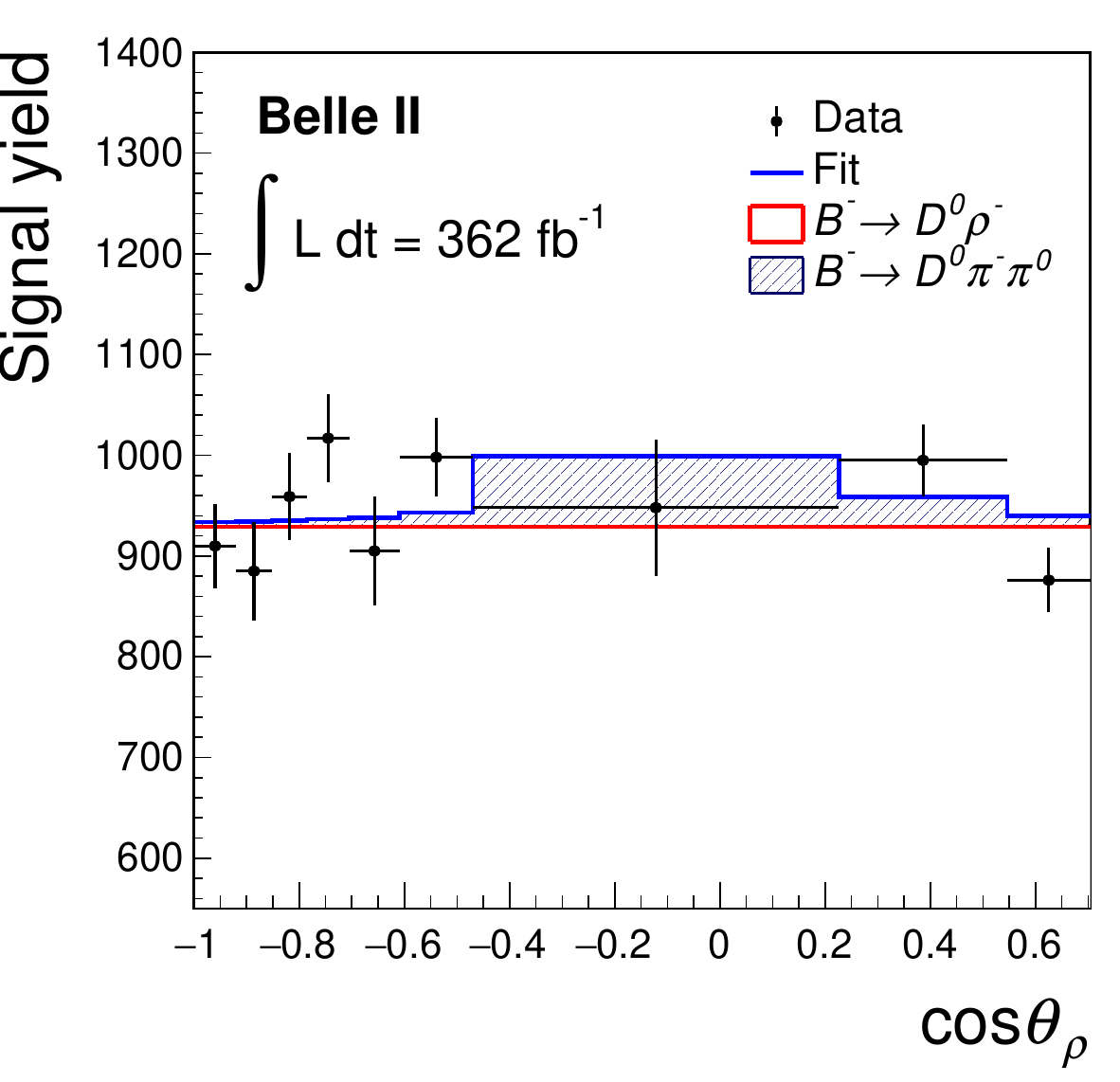}
     \includegraphics[width=0.80\linewidth]{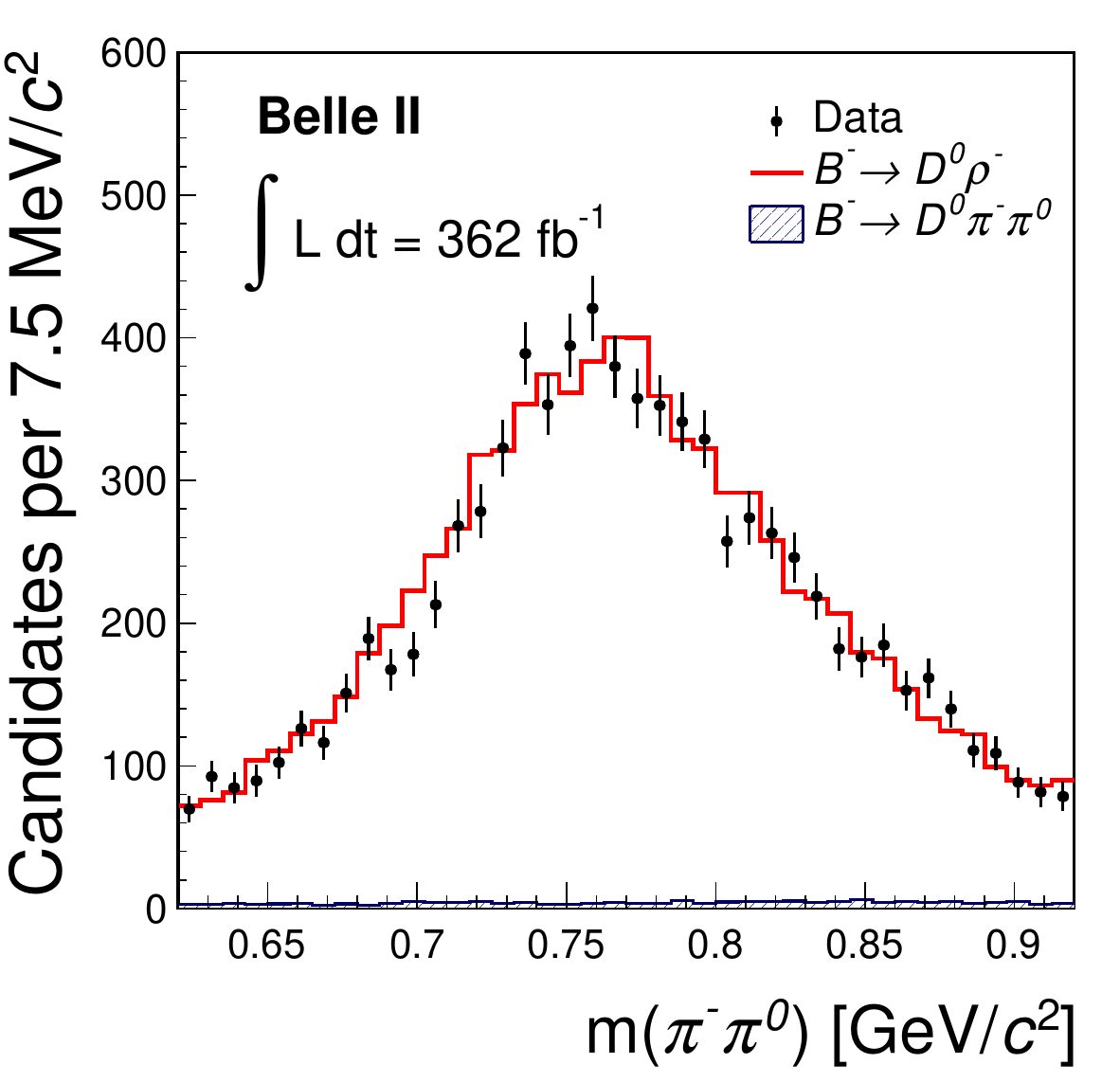}
    \caption{Top: background-subtracted $\cos\theta_{\rho}$ distribution with fit projection overlaid. The horizontal bar on the markers represents the $\cos\theta_{\rho}$ interval. The $y$-axis scale is zero-suppressed to show the small \BDpipi contribution. Bottom: background-subtracted distribution of the $\pim\piz$ invariant mass, with overlaid simulated data reflecting the \BDrho and \BDpipi proportions from the $\cos\theta_{\rho}$ fit. }
    \label{fig:results}
\end{figure}

The background-subtracted $\cos\theta_{\rho}$ distribution is shown in Fig.~\ref{fig:results} (top). We carry out a $\chi^2$ fit to this distribution to extract the \BDrho signal yield. Since we adopt an interval scheme with equiprobable \BDrho yields,  we use a uniform probability density function to model \BDrho candidates and a peaking template obtained from simulation to model \BDpipi candidates. We find $8360\pm180$ \BDrho decays and a $(1.9\pm1.8)\%$ fraction of \BDpipi decays. The $p$ value of the fit is 0.074. Figure~\ref{fig:results} (bottom) shows the background-subtracted distribution of the $\pim\piz$ invariant mass, overlaid with the simulated distribution using the fractions of \BDrho and \BDpipi decays resulting from the fit. Background is subtracted using the sPlot method~\cite{Pivk:2004ty} with per-candidate weights calculated from the $\Delta E$ fits in the nine $\cos\theta_{\rho}$ intervals. Simulation reproduces the data distribution well. We check the stability of the obtained results by separately fitting the candidates with $\pim\piz$ invariant masses above and below 770\mevcc. We obtain compatible $\cos\theta_\rho$ distributions, with fractions of nonresonant \BDpipi decays again consistent with zero and branching fractions of \BDrho decays that agree within half a standard deviation considering only the statistical uncertainties. These results indicate that interference effects between the two amplitudes are  negligible. 

\begin{table}[t]
  \caption{Summary of the fractional systematic uncertainties. The total is the sum in quadrature of the individual contributions.}
  \label{tab:syst_br}
  \centering
  \begin{tabular}{ l  c  } 
\hline
  \hline
Source  & Fractional uncertainty (\%)\\
\hline
$N_{\BB}$ & 1.5 \\
 $f^{+-}$ & 2.4 \\
$\mathcal{B}_{\rm sub}$ & 0.8 \\
 Fit modeling & 1.7 \\ 
 $\piz$ efficiency  &  3.7 \\
 Particle-identification efficiency   & 0.6 \\
Continuum-suppression efficiency & 1.5 \\
Tracking efficiency& 0.7 \\
 \hline
 Total & 5.3 \\
\hline
  \hline
  \end{tabular}
\end{table}

The branching fraction of the \BDrho decay is given by 
\begin{equation}
 \mathcal{B}(\BDrho) = \frac{N}{2\, N_{\BB} \, f^{+-}\,  \varepsilon \, \mathcal{B}_{\rm sub}}\,,
\end{equation}
where $N$ is the \BDrho yield, $N_{\BB}=(387 \pm 6)\times 10^6$ the number of \BB pairs in the sample, $f^{+-}=0.516 \pm 0.012$ the branching fraction of the decay $\FourS \to B^+B^-$~\cite{f+-/f00}, $\varepsilon=(5.71 \pm 0.03)\%$ the signal efficiency, and $\mathcal{B}_{\rm sub}= (3.90 \pm 0.03)\%$ the product of the branching fractions of the subdecays $D^0\to K^-\pi^+$ and $\pi^0\to \gamma\gamma$~\cite{PDG_2022}. The result is
\begin{equation}
\mathcal{B}(\BDrho) = (\BFCentValue \pm \BFStatErr \pm \BFSystErr)\%\,,
\end{equation}
where the first uncertainty is statistical and the second systematic. This value already includes the simulation corrections for the efficiency described below. 

Contributions to the systematic uncertainty are reported in Table~\ref{tab:syst_br}. The uncertainties on the number $N_{\BB}$ of $\BB$ pairs, on the branching fraction $f^{+-}$, and on the subdecay branching fraction $\mathcal{B}_{\rm sub}$ contribute systematic uncertainties of 1.5\%, 2.4\%, and 0.8\%, respectively. 

We calculate systematic uncertainties associated with the choice of the $\Delta E$ fit model, for each component, by drawing one thousand samples of pseudodata from an alternative model and fitting them with the model used in the analysis (nominal model). The alternative models consist of different probability density functions that fit simulated data with the same quality as the nominal models. In addition, we generate samples with variations of  parameters that are fixed to the values found from the fit to simulated data and fit them with the default model, to mimic potential data-simulation discrepancies. We assign as a systematic uncertainty the observed average residual, {\it i.e.}, the difference between the average fit value and the true value, of the signal \BDrho yield.  
We sum in quadrature the systematic uncertainties related to the model choices and to the fixed parameters to obtain the systematic uncertainty associated with the fit modeling reported in Table~\ref{tab:syst_br}. 

We validate the background composition by inspecting several distributions in two sidebands of the diphoton invariant mass (110\,--120 and 145\,--160\mevcc), which are enriched in self-cross-feed and \BB background. A 10\% relative uncertainty is assigned to the fraction of self-cross-feed determined from simulated data in each $\cos\theta_\rho$ interval. This uncertainty is calculated in sideband data by comparing the simulated and experimental distributions of the $\pi^0$ momentum, which is the kinematic variable most sensitive to variation of the self-cross-feed fraction. Since this fraction is Gaussian constrained in the $\Delta E$ fit, the corresponding systematic uncertainty is already included in the statistical uncertainty on the signal yields. 

The signal efficiency is calculated from simulated data; the quoted $0.03\%$ uncertainty is statistical. Control data are used to validate the efficiency  and to correct for data-simulation differences.  We propagate uncertainties of all efficiency corrections as systematic uncertainties on the branching fraction. 

The correction for the $\piz$-reconstruction efficiency is obtained by measuring the ratio of the yields of the decays $\Dstarp \to \Dz (\to \Km \pip \piz)\pip$ and $ \Dstarp \to \Dz (\to \Km \pip) \pip$, scaled by the inverse of the branching-fraction ratio.  We measure the yield ratio in experimental and simulated data to determine a per-candidate correction as a function of the momentum and polar angle of the $\piz$ candidate. The corrections span a range between $0.7$ and $1.1$ with an average of $1.011\pm0.037$. The uncertainty is dominated by that on the ratio of $D^0$-decay branching fractions~\cite{PDG_2022}.

The correction for the efficiency of the kaon selection is obtained using an abundant control sample of $\Dstarp \to \Dz (\to \Km \pip) \pip$ decays. We measure the data-to-simulation efficiency ratio from the control sample and scale the signal efficiency using per-candidate corrections as a function of the momentum and polar angle of the kaon candidate. The corrections span a range between $0.7$ and $1.2$ with an average of $1.009 \pm 0.005$. The uncertainty is dominated by that on the background subtraction of the control channels.

The efficiency of the continuum-suppression requirement is validated using   \BDpiKpi decays. We use a signal-enriched region (signal purity of 95\%) and a continuum-background-dominated region (continuum-background fraction of 83\%) to compare distributions of  the BDT input variables and BDT output between experimental and simulated data. No significant discrepancy is found. The ratio of efficiency of the continuum-suppression requirement in experimental and simulated control data is $0.987 \pm 0.015$, where the uncertainty is statistical. We use this ratio to correct the signal efficiency and propagate the corresponding  uncertainty to the signal branching fraction.

The tracking efficiency is validated using $\epem \to \tautau$ events, where one $\tau$ decays leptonically, $\taup \to \ell^+ \nu_{\ell} \nub_{\tau}$ with $\ell = e, \mu$, and the other hadronically, $\taum \to \pim \pi^+\pi^-\nu_{\tau}$. The efficiency to reconstruct a track is found to be the same for experimental and simulated data with an uncertainty of $0.24\%$. No correction is applied to the signal efficiency. 

Finally, we validate the overall efficiency and its corrections by measuring the branching fraction of the \BDpiKpipiz decay, selected  as the signal (except for requirements on $\pim\piz$ invariant-mass and $\cos\theta_\rho$).  Although the final-state particles are the same, the kinematics of the control decay differs from that of the signal decay. This results in a selection efficiency for the control channel of $(2.98 \pm 0.03)$\%, where the difference with that of the signal is mainly due to the charged pion accompanying the \Dz candidate in the $B^-\to\Dz\pim$ decay.  However, for the other particles, the \BDpiKpipiz decay covers most of the relevant \BDrho phase-space. In particular, the neutral pion from the \BDpiKpipiz decay has a softer momentum distribution that peaks in the region where the efficiency corrections are larger.  As in the signal analysis, the control sample is categorized by three components: signal \BDpiKpipiz (75\% fraction), \BB background (21\%), and self-cross-feed (3\%). Continuum background is negligible. 
To determine the signal yield, we use a maximum likelihood fit to the unbinned $\Delta E$ distribution using probability density functions similar to those described for the signal $\Delta E$ fit. We obtain $7700 \pm 90$ \BDpiKpipiz decays, which yields a branching fraction of $(0.454 \pm 0.022)$\% in agreement with the world average value, $(0.461 \pm 0.010)$\%~\cite{PDG_2022}. The result includes the efficiency corrections and their uncertainties, as well as those due to the number of \BB pairs, in $f^{+-}$, and in the subdecay branching fractions. 


Using our result, we update the value of $R$ and $\cos\delta$ in Eqs~(\ref{eq:R}) and~(\ref{eq:delta}). Taking the branching fractions $\mathcal{B}(\Bbar^0 \to D^+ \rho^-)=(0.76 \pm 0.12)\%$ and $\mathcal{B}(\Bbar^0 \to D^0 \rho^0)=(0.0321 \pm 0.0021)\%$, and the ratio of lifetimes $\tau_+/\tau_0=1.076 \pm 0.004$~\cite{PDG_2022}, we obtain $R = 0.93^{+0.11}_{-0.12}$ and $\cos\delta = 0.919^{+0.012}_{-0.009}$, which agree with and are significantly more precise than previous determinations~\cite{LHCb:2015klp}. These results confirm expectations from factorization in the heavy-quark limit. 

In conclusion, we measure the branching fraction of the \BDrho decay using \epem-collision data collected by the Belle II detector at the $\FourS$ resonance and containing 387 million \BB meson pairs. The result is $(\BFCentValue \pm \BFStatErr\stat \pm \BFSystErr \syst)\%$, in agreement with previous determinations. Our measurement improves  the fractional precision of the world average~\cite{PDG_2022} by more than a factor of two and will also significantly ameliorate the calibration factor of the Belle II hadronic-tagging algorithm. 

\input{acknowledgements}


\bibliography{belle2}
\bibliographystyle{apsrev4-1}

\clearpage
\onecolumngrid

\end{document}

%% file: authors.tex
  \author{I.~Adachi\,\orcidlink{0000-0003-2287-0173}} 
  \author{L.~Aggarwal\,\orcidlink{0000-0002-0909-7537}} 
  \author{H.~Aihara\,\orcidlink{0000-0002-1907-5964}} 
  \author{N.~Akopov\,\orcidlink{0000-0002-4425-2096}} 
  \author{A.~Aloisio\,\orcidlink{0000-0002-3883-6693}} 
  \author{N.~Anh~Ky\,\orcidlink{0000-0003-0471-197X}} 
  \author{D.~M.~Asner\,\orcidlink{0000-0002-1586-5790}} 
  \author{H.~Atmacan\,\orcidlink{0000-0003-2435-501X}} 
  \author{V.~Aushev\,\orcidlink{0000-0002-8588-5308}} 
  \author{M.~Aversano\,\orcidlink{0000-0001-9980-0953}} 
% ...
\collaboration{The Belle II Collaboration}

%% file: acknowledgements.tex
This work, based on data collected using the Belle II detector, which was built and commissioned prior to March 2019,
was supported by
Higher Education and Science Committee of the Republic of Armenia Grant No.~23LCG-1C011;
Australian Research Council and Research Grants
No.~DP200101792, 
No.~DP210101900, 
No.~DP210102831, 
No.~DE220100462, 
No.~LE210100098, 
and
No.~LE230100085; 
Austrian Federal Ministry of Education, Science and Research,
Austrian Science Fund
No.~P~31361-N36
and
No.~J4625-N,
and
Horizon 2020 ERC Starting Grant No.~947006 ``InterLeptons'';
Natural Sciences and Engineering Research Council of Canada, Compute Canada and CANARIE;
National Key R\&D Program of China under Contract No.~2022YFA1601903,
National Natural Science Foundation of China and Research Grants
No.~11575017,
No.~11761141009,
No.~11705209,
No.~11975076,
No.~12135005,
No.~12150004,
No.~12161141008,
and
No.~12175041,
and Shandong Provincial Natural Science Foundation Project~ZR2022JQ02;
the Czech Science Foundation Grant No.~22-18469S;
European Research Council, Seventh Framework PIEF-GA-2013-622527,
Horizon 2020 ERC-Advanced Grants No.~267104 and No.~884719,
Horizon 2020 ERC-Consolidator Grant No.~819127,
Horizon 2020 Marie Sklodowska-Curie Grant Agreement No.~700525 ``NIOBE''
and
No.~101026516,
and
Horizon 2020 Marie Sklodowska-Curie RISE project JENNIFER2 Grant Agreement No.~822070 (European grants);
L'Institut National de Physique Nucl\'{e}aire et de Physique des Particules (IN2P3) du CNRS
and
L'Agence Nationale de la Recherche (ANR) under grant ANR-21-CE31-0009 (France);
BMBF, DFG, HGF, MPG, and AvH Foundation (Germany);
Department of Atomic Energy under Project Identification No.~RTI 4002,
Department of Science and Technology,
and
UPES SEED funding programs
No.~UPES/R\&D-SEED-INFRA/17052023/01 and
No.~UPES/R\&D-SOE/20062022/06 (India);
Israel Science Foundation Grant No.~2476/17,
U.S.-Israel Binational Science Foundation Grant No.~2016113, and
Israel Ministry of Science Grant No.~3-16543;
Istituto Nazionale di Fisica Nucleare and the Research Grants BELLE2;
Japan Society for the Promotion of Science, Grant-in-Aid for Scientific Research Grants
No.~16H03968,
No.~16H03993,
No.~16H06492,
No.~16K05323,
No.~17H01133,
No.~17H05405,
No.~18K03621,
No.~18H03710,
No.~18H05226,
No.~19H00682, 
No.~20H05850,
No.~20H05858,
No.~22H00144,
No.~22K14056,
No.~22K21347,
No.~23H05433,
No.~26220706,
and
No.~26400255,
and
the Ministry of Education, Culture, Sports, Science, and Technology (MEXT) of Japan;  
National Research Foundation (NRF) of Korea Grants
No.~2016R1\-D1A1B\-02012900,
No.~2018R1\-A2B\-3003643,
No.~2018R1\-A6A1A\-06024970,
No.~2019R1\-I1A3A\-01058933,
No.~2021R1\-A6A1A\-03043957,
No.~2021R1\-F1A\-1060423,
No.~2021R1\-F1A\-1064008,
No.~2022R1\-A2C\-1003993,
and
No.~RS-2022-00197659,
Radiation Science Research Institute,
Foreign Large-Size Research Facility Application Supporting project,
the Global Science Experimental Data Hub Center of the Korea Institute of Science and Technology Information
and
KREONET/GLORIAD;
Universiti Malaya RU grant, Akademi Sains Malaysia, and Ministry of Education Malaysia;
Frontiers of Science Program Contracts
No.~FOINS-296,
No.~CB-221329,
No.~CB-236394,
No.~CB-254409,
and
No.~CB-180023, and SEP-CINVESTAV Research Grant No.~237 (Mexico);
the Polish Ministry of Science and Higher Education and the National Science Center;
the Ministry of Science and Higher Education of the Russian Federation
and
the HSE University Basic Research Program, Moscow;
University of Tabuk Research Grants
No.~S-0256-1438 and No.~S-0280-1439 (Saudi Arabia);
Slovenian Research Agency and Research Grants
No.~J1-9124
and
No.~P1-0135;
Agencia Estatal de Investigacion, Spain
Grant No.~RYC2020-029875-I
and
Generalitat Valenciana, Spain
Grant No.~CIDEGENT/2018/020;
National Science and Technology Council,
and
Ministry of Education (Taiwan);
Thailand Center of Excellence in Physics;
TUBITAK ULAKBIM (Turkey);
National Research Foundation of Ukraine, Project No.~2020.02/0257,
and
Ministry of Education and Science of Ukraine;
the U.S. National Science Foundation and Research Grants
No.~PHY-1913789 
and
No.~PHY-2111604, 
and the U.S. Department of Energy and Research Awards
No.~DE-AC06-76RLO1830, 
No.~DE-SC0007983, 
No.~DE-SC0009824, 
No.~DE-SC0009973, 
No.~DE-SC0010007, 
No.~DE-SC0010073, 
No.~DE-SC0010118, 
No.~DE-SC0010504, 
No.~DE-SC0011784, 
No.~DE-SC0012704, 
No.~DE-SC0019230, 
No.~DE-SC0021274, 
No.~DE-SC0021616, 
No.~DE-SC0022350, 
No.~DE-SC0023470; 
and
the Vietnam Academy of Science and Technology (VAST) under Grants
No.~NVCC.05.12/22-23
and
No.~DL0000.02/24-25.

These acknowledgements are not to be interpreted as an endorsement of any statement made
by any of our institutes, funding agencies, governments, or their representatives.

We thank the SuperKEKB team for delivering high-luminosity collisions;
the KEK cryogenics group for the efficient operation of the detector solenoid magnet;
the KEK Computer Research Center for on-site computing support; the NII for SINET6 network support;
and the raw-data centers hosted by BNL, DESY, GridKa, IN2P3, INFN, 
and the University of Victoria.